\documentclass[12pt]{article}
\usepackage{graphicx,amsmath,amssymb,epsfig}

\newlength{\dinwidth}
\newlength{\dinmargin}
\def\lapproxeq{\lower .7ex\hbox{$\;\stackrel{\textstyle
<}{\sim}\;$}}
\def\gapproxeq{\lower .7ex\hbox{$\;\stackrel{\textstyle
>}{\sim}\;$}}
\def\gtrsim{\lower .7ex\hbox{$\;\stackrel{\textstyle
>}{\sim}\;$}}
\def\lesim{\lower .7ex\hbox{$\;\stackrel{\textstyle
<}{\sim}\;$}}

\def\be{\begin{equation}}
\def\ee{\end{equation}}
\def\bea{\begin{eqnarray}}
\def\eea{\end{eqnarray}}
\def\funp{{I\!\!P}}

\def\GeV{\rm GeV}

\begin{document}
\begin{flushright}
IPPP/08/15 \\
DCPT/08/30 \\
 March 2008 \\

\end{flushright}

\vspace*{0.5cm}

\begin{center}
{\Large \bf Testing predictions for central exclusive processes in the early LHC runs\footnote{
Presented 
by V.A.~Khoze
at the 22nd Workshop Les Rencontres de Physique de la Vall\'{e}e d'Aoste, La Thuile, 
Aosta Valley, 24 February- March 1, 2008.}} \\   

\vspace*{1cm}
\textsc{V.A.~Khoze$^{a,b}$, A.D. Martin$^a$ and M.G. Ryskin$^{a,b}$} \\

\vspace*{0.5cm}
$^a$ Department of Physics and Institute for
Particle Physics Phenomenology, \\
University of Durham, DH1 3LE, UK \\
$^b$ Petersburg Nuclear Physics Institute, Gatchina,
St.~Petersburg, 188300, Russia \\
\end{center}

\vspace*{0.5cm}

\begin{abstract}
We show that the early LHC measurements can provide
crucial checks of the different components of the 
formalism used to predict the cross sections of central exclusive processes.
\end{abstract}

\newpage
\section{Introduction}
The benefits of using forward proton detectors
as a tool to study Standard Model (SM) and New Physics at the LHC
have been fully appreciated only recently, see for
instance, \cite{INC}~-~\cite{KMRlt}.
The measurements of central exclusive production (CEP) 
is a prime target of the FP420
project \cite{LOI}, which aims at the installation of
forward detectors in the LHC tunnel 420 m from the
interaction points of the ATLAS and CMS experiments.
The combined detection of both outgoing protons and the centrally
produced system gives access to a rich programme
of studies in QCD, electroweak, Higgs and BSM physics. Importantly,
 these measurements will 
 provide valuable information on the Higgs sector of MSSM and other popular
BSM scenarios, see \cite{KKMRext}~-~\cite{fghpp}. 
In particular, the CEP
process allows for the
  unique possibility to study directly the coupling of
 Higgs-like bosons to bottom quarks
\cite{INC,KMRmm}.

\begin{figure}
\begin{center}
\includegraphics[height=6cm]{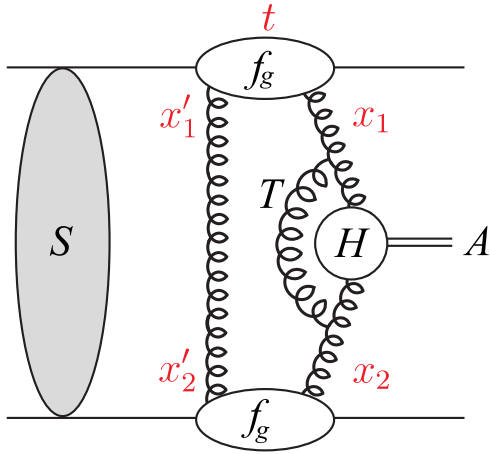}
\caption{A symbolic diagram for the central exclusive production of a system $A$.}
\label{fig:parts}
\end{center}
\end{figure}

The theoretical formalism \cite{KMR}~-~\cite{KMRnewsoft}
 for the description of 
a CEP process contains quite distinct parts, shown symbolically
in Fig.~\ref{fig:parts}.
We first have to calculate the $gg \to A$ subprocess, $H$, convoluted with the gluon distributions $f_g$. Next we must account for the QCD corrections which reflect the absence of additional QCD radiation in the hard subprocess -- that is, for the Sudakov factor $T$. Finally, we must enter soft physics to calculate the survival probability $S^2$ of the rapidity gaps.

The uncertainties of the  CEP predictions
are potentially not small. Therefore,
 it is important to perform checks using processes  that will be accessible in the first  runs of the LHC
\cite{KMRearly} 
with integrated luminosities in the range 100 ${\rm pb}^{-1}$ to 1 ${\rm fb}^{-1}$.
In \cite{KMRearly} we
identified reactions where the different ingredients of the formalism used to calculate CEP 
could be tested experimentally.
We first consider measurements which do not rely on proton tagging and 
can be performed through the detection of rapidity gaps. 

The main uncertainties of the CEP predictions  are associated with 
\begin{itemize}
\item [(i)] the probability $S^2$ that additional secondaries will not populate the gaps; 

\item [(ii)] the probability to find the appropriate gluons, that are given by generalized, unintegrated distributions $f_g(x,x',Q_t^2)$;
\item [(iii)] the higher order QCD corrections to the hard subprocess, where the most important is the  Sudakov suppression; 
\item [(iv)] the so-called semi-enhanced absorptive corrections (see \cite{KKMR,bbkm}) and other effects, which may violate the soft-hard factorization.
\end{itemize}


\section{Gap survival probability $S^2$}

As a rule, the gap survival probability is calculated within a multichannel eikonal approach \cite{GLMrev}. 
The probability $S^2$ of elastic $pp$ rescattering, 
shown symbolically  by $S$ in 
Fig.~\ref{fig:parts} 
can be evaluated in a model independent way once  the 
elastic cross section $d\sigma_{\rm el}/dt$ is measured at the LHC. However, there may be some excited states  between the blob $S$ and the   amplitude on the right-hand-side of Fig.~\ref{fig:parts}.
  The presence of such states enlarges the absorptive correction.  In order to experimentally check the role of this effect, we need to consider a process with a bare cross section that can be reliably calculated. Good candidates are the production of $W$ or $Z$ bosons with rapidity gaps.
\begin{figure}
\begin{center}
\includegraphics[height=6cm]{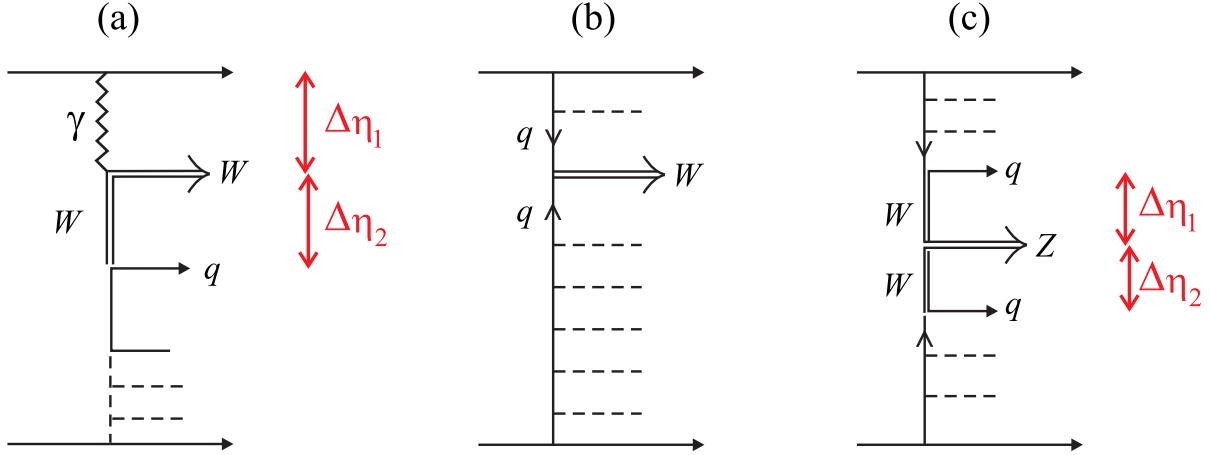}
\caption{Diagrams for (a) $W$ production with 2 rapidity gaps, (b) inclusive $W$ production, and (c) $Z$ production with 2 rapidity gaps.}
\label{fig:WZ}
\end{center}
\end{figure}
In the case of `$W$+gaps' production the main contribution comes from the diagram shown
 in Fig.~\ref{fig:WZ}(a) 
\cite{KMRphoton}. One gap, $\Delta \eta_1$, is associated with photon exchange, while the other, $\Delta \eta_2$, is associated with the $W$. The cross section is proportional to the quark distribution  at a large scale and not too small $x$, where the uncertainties of the parton densities are small.
 To select these events in the early LHC runs, we can use the rapidity gap  trigger combined with a high $p_t$  lepton or jet trigger.

\begin{figure} 
\begin{center}
\includegraphics[height=9cm]{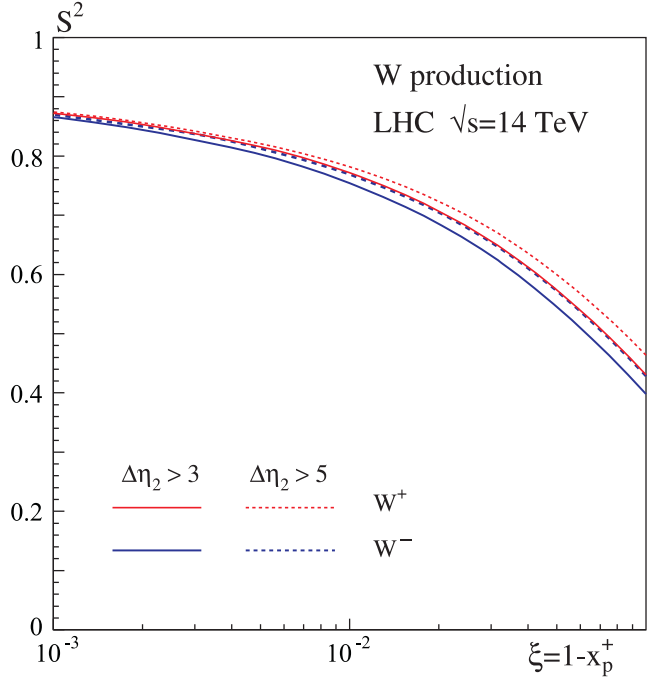}
\caption{The rapidity gap survival factor $S^2$ as a function of $\xi$ calculated using the global soft model of \cite{KMRsoft}, assuming that the valence (sea) quarks are associated with the weak (strong) absorptive components. The small spread of the predictions arising from the different partonic content of the diffractive eigenstates means that $W$+gaps events offer a meaningful test of the $S^2$ factor.} 
\label{fig:W3}
\end{center}
\end{figure}

An important point here is that
 the minimum value of $|t|$ of the photon,
$|t_{\rm min}|~\simeq~\frac{m_N^2 \xi^2}{1-\xi}$,
is not negligible. 
Note that the momentum fraction $x_p=1-\xi$ associated with the upper proton can be measured
by summing the momentum fractions of the outgoing $W$ and the hadrons observed in the calorimeters.
As illustrated in Fig.~\ref{fig:W3}
the rescattering reduces the cross section by the factor $S^2$. 
The curves in Fig.~\ref{fig:W3} were calculated within the scenario where the valence quark is
allocated 
to the component with the smallest absorption and the sea quark to the 
absorptive component with largest cross section. 
Since the valence quark contribution is more important for $W^+$ production and for the configuration with the largest gap size $\Delta\eta_2$, the expected gap survival factor $S^2$ is found to be larger.
In the first LHC data runs the ratio ($W$+gaps/$W$ inclusive) will be measured first.
This measurement is a useful check of the models for soft rescattering.

A good way 
to study the low impact parameter ($b_t$) region 
 is to observe $Z$ boson production via $WW$ fusion, see Fig.~\ref{fig:WZ}(c). Here, both of the rapidity gaps originate from heavy boson exchange, and the corresponding $b_t$ region is similar to that for exclusive Higgs production. The expected $Z$+gaps cross section is of the order of 0.2 pb, and $S^2$=0.3 for $\Delta \eta_{1,2} > 3$ and for quark jets with $E_T>50$ GeV \cite{pw}.

One problem is that even with the $E_T>50$ GeV cut, the QCD background arising from  QCD $b\bar{b}$ central exclusive production is comparable to the electroweak $q\bar{q} \to Z+2$ jet signal. Therefore, we should concentrate on the leptonic decay modes of the $Z$ boson, which results in a smaller event rate
\footnote{Note that in the recent study \cite{nikit} it was demonstrated that the so-called Track Counting Veto (TCV) is robust for selection of the central
rapidity gap events in vector-boson fusion 
searches at CMS.}.


When the forward proton detectors become operational we can do more. Both the longitudinal and transverse momentum of the protons can be measured, and we can study the $k_t$ behaviour of the cross section sections  and scan the proton opacity \cite{KMRphoton}.

\section{Generalized, unintegrated gluon distribution $f_g$}
The cross section for the CEP of a system $A$ 
essentially has the form \cite{KMR}
\begin{equation}
\sigma(pp \to p+A+p) ~\simeq ~\frac{S^2}{b^2} \left|\frac{\pi}{8} \int\frac{dQ^2_t}{Q^4_t}\: f_g(x_1, x_1', Q_t^2, \mu^2)f_g(x_2,x_2',Q_t^2,\mu^2)~ \right| ^2~\hat{\sigma}(gg \to A).
\label{eq:M}
\end{equation}
Here the factor $1/b^2$ arises from the integration over the proton transverse momentum.  Also, $f_g$ denotes the generalized unintegrated gluon distribution in the limit of $p_t \to 0$. 
The distribution $f_g$ has not yet been measured explicitly. However, in our case it can be obtained from the conventional diagonal gluon distribution, $g$, known from the global parton analyses, see \cite{KMR,KMRearly} for details.
The main uncertainty here comes from the lack of knowledge of the integrated gluon distribution $g(x,Q_t^2)$ at low $x$ and small scales. For example, taking $Q_t^2=4~\GeV^2$ we find that a variety of recent MRST \cite{mstw} and CTEQ \cite{cteq} global analyses give a spread of
$xg~=~(3-3.8)~~{\rm for}~~x=10^{-2}~~~~{\rm and}~~~~xg~=~(3.4-4.5)~~{\rm for}~~x=10^{-3}$.
These are big uncertainties bearing in mind that the CEP cross section 
 depends on $(xg)^4$.
   
\begin{figure} [t]
\begin{center}
\includegraphics[height=5cm]{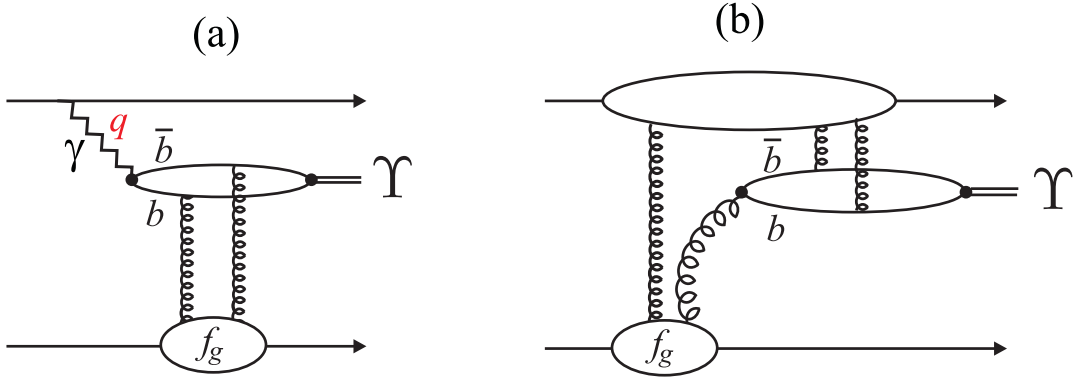}
\caption{Exclusive $\Upsilon$ production via (a) photon exchange, and (b) via odderon exchange.}
\label{fig:upsilon}
\end{center}
\end{figure}

To reduce the uncertainty associated with $f_g$ we can measure exclusive $\Upsilon$ production
\footnote{
A feasibility study of the $\gamma p \to \Upsilon p$ measurement performed by CMS \cite{YCMS}
looks quite encouraging.}.
 The process is shown in Fig.~\ref{fig:upsilon}(a).  The cross section for $\gamma p \to \Upsilon p$ \cite{mrt} is given in terms of exactly the same  unintegrated gluon distribution $f_g$ that occurs in Fig.~\ref{fig:parts}.

There may be competition between production via photon exchange, Fig.~\ref{fig:upsilon}(a), and via odderon exchange,
 see Fig.~\ref{fig:upsilon}(b). To date, odderon exchange has not been observed. On the other hand, 
a lowest-order  calculation indicates that the odderon process (b) may be comparable to the photon-initiated process (a) (for example, \cite{bmsc}). If the upper proton is tagged, it will be straightforward to separate the two mechanisms
since odderon production has no $1/q^2$.
singularity characteristic of the photon. 

The expression for $\sigma(\gamma p \to \Upsilon p) \propto f_g^2$ is given in \cite{mrt}. 
In order to use this process to constrain the gluon distribution
it would be preferable to tag the lower proton.

\section{Three-jet events as a probe of the Sudakov factor}

Traditionally, the search for the exclusive dijet signal at the Tevatron, $p\bar{p} \to p+jj+\bar{p}$, is performed \cite {CDFjj} by plotting the cross section in terms of the variable
$R_{jj}=M_{jj}/M_A$, where $M_A$ is the mass of the whole central system.
The $R_{jj}$  distribution is strongly smeared out by
QCD bremsstrahlung, hadronization, the jet searching algorithm and other experimental effects \cite {CDFjj,KMRrj}.
\begin{figure}
\begin{center}
\includegraphics[height=7cm]{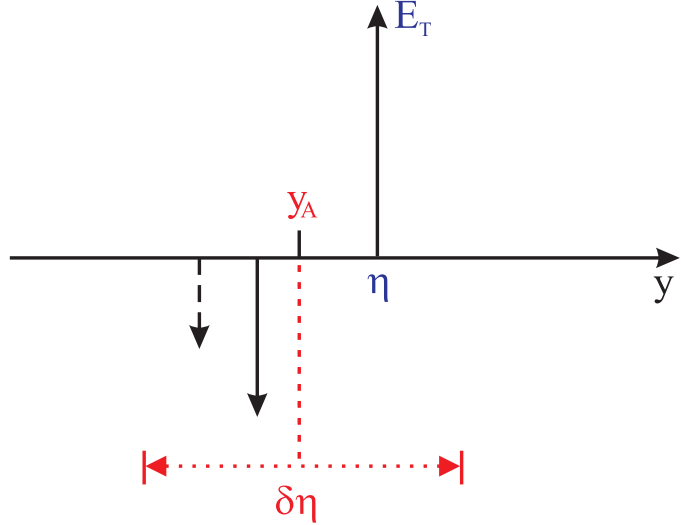}
\caption{The rapidities of the three jets in the central system. Note that the rapidity $y_A$ of the whole central system does not necessarily occur at $y=0$. The rapidity
interval containing the three jets is denoted by $\delta\eta$, outside of which there is no hadronic
activity.\label{fig:2}}
\end{center}
\end{figure}
To weaken the role of  smearing it was proposed in Ref.~\cite{KMRrj} 
to study the  dijet distribution in terms of a new variable
\be
R_j~=~2E_T ~({\rm cosh}~\eta^*)/M_A~,
\label{eq:j}
\ee
where only the transverse energy and the rapidity $\eta$ of the jet with
the {\it largest} $E_T$ are used in
the numerator.  Here $\eta^* = \eta -y_A$ where $y_A$ is the rapidity of the whole central system.
Clearly
the jet with the largest $E_T$ is less affected by hadronization, final parton radiation etc.  
At leading order, it is sufficient to consider
the emission of a third  jet, as shown in Fig.~\ref{fig:2}, where we take all three jets to lie in a specified rapidity interval $\delta\eta$. 

The cross section $d\sigma/dR_j$, as a function of $R_j$, for the exclusive production of a
pair of  high $E_T$ dijets accompanied by a third  jet was calculated and discussed in \cite{KMRrj,KMRearly}. 
It was shown that 
 the measurements of the exclusive two- and three-jet cross sections {\it as a function of $E_T$} of the highest jet allow a detailed check of the Sudakov physics; with much more information coming from the observation of the $\delta\eta$ dependence. 

A clear way to observe the effect of the Sudakov suppression is just to study the $E_T$ dependence of exclusive dijet production. On dimensional grounds we would expect $d\sigma/dE_T^2 \propto 1/E_T^4$. This behaviour is modified by the anomalous dimension of the gluon and by a stronger Sudakov suppression with increasing $E_T$. Already the existing CDF  dijet data \cite{CDFjj} exclude predictions which omit the Sudakov effect.
\section{Soft-hard factorization: enhanced absorptive effects}
\begin{figure} 
\begin{center}
\includegraphics[height=6cm]{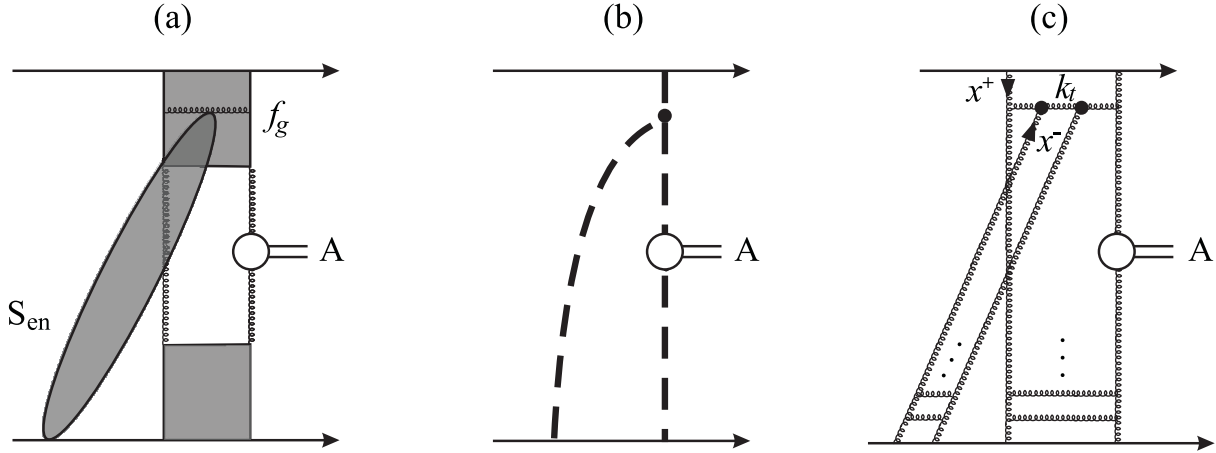}
\caption{(a) A typical enhanced diagram, where the shaded boxes symbolically denote $f_g$, and the soft rescattering is on an intermediate parton, giving rise to a gap survival factor $S_{\rm en}$; (b) and (c) are the Reggeon and QCD representations, respectively.}
\label{fig:enh}
\end{center}
\end{figure}
The soft-hard factorization implied by Fig.~\ref{fig:parts}
could be violated by the so-called enhanced Reggeon diagrams, caused by the rescattering of an intermediate parton generated in the evolution of $f_g$. Such a diagram is shown in Fig.~\ref{fig:enh}(a).

The contribution of the first Pomeron loop diagram, Fig.~\ref{fig:enh}(b) was calculated in pQCD in Ref.~\cite{bbkm}. A typical  diagram is shown in Fig.~\ref{fig:enh}(c). For LHC energies it was found that the probability of such rescattering may be numerically large.
The reason is that the gluon density grows in the low $x$ region and, for low $k_t$ partons, approaches the saturation limit. However, as shown in \cite{KMRearly},     
the enhanced diagram should affect mainly the very beginning of the QCD evolution -- the region that cannot be described perturbatively and which, in the calculations of \cite{KMRsoft,KMRnewsoft},
is already included phenomenologically.

Experimentally,  
we can study the role of semi-enhanced absorption  by measuring the
 ratio $R$ of diffractive events for $W$ (or $\Upsilon$ or dijet) production as compared to the inclusive process \cite{KMRearly}.
 That is
\be 
R~~=~~\frac{{\rm no.~of}~ (A+{\rm gap)~ events}}{{\rm no.~of~ (inclusive}~A)~{\rm events}}~~=~~
\frac{a^{\rm diff}(x_\funp ,\beta,\mu^2)}{a^{\rm incl}(x=\beta x_\funp,\mu^2)}~\langle S^2S^2_{\rm en}\rangle_{{\rm over}~b_t},
\label{eq:Ren}
\ee
where $a^{\rm incl}$ and $a^{\rm diff}$ are the parton densities determined from the global analyses of inclusive and diffractive DIS data, respectively.
 For $W$ or $\mu^+\mu^-$  production the parton densities $a$ are quark distributions, whereas for dijet or $\Upsilon$ they are mainly gluon densities. 

\begin{figure} 
\begin{center}
\includegraphics[height=15cm]{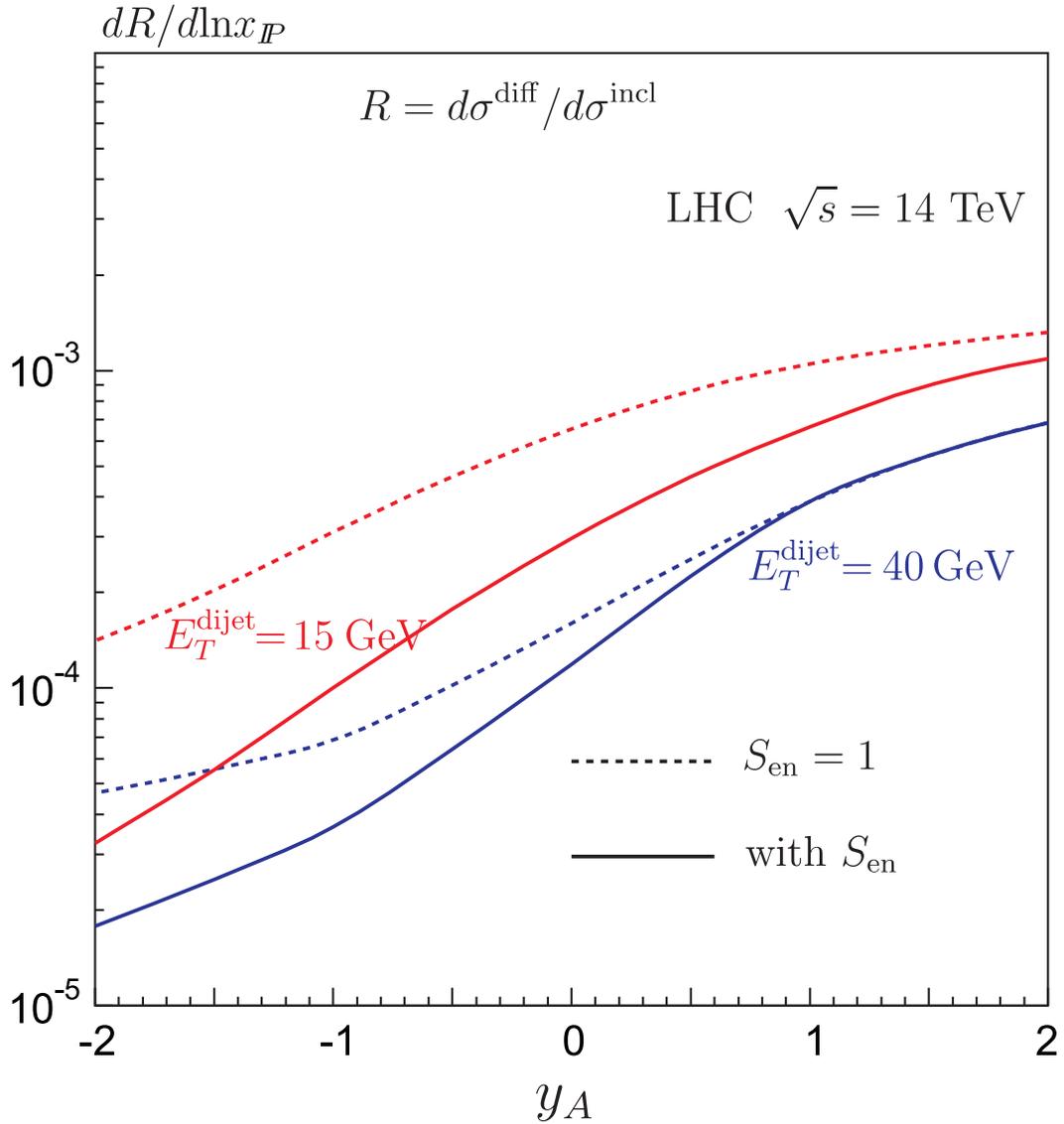}
\caption{The predictions of the ratio $R$ of (\ref{eq:Ren}) for the production of a pair of high $E_T$ jets with (continuous curves) and without (dashed curves) enhanced soft rescattering on intermediate partons.}
\label{fig:upsd}
\end{center}
\end{figure}

Experimentally,
we can observe a double distribution $d^2 \sigma^{\rm diff}/dx_\funp dy_A$, and form the ratio $R$ using the inclusive cross section, $d\sigma^{\rm incl}/dy_A$.
 If we neglect the enhanced absorption, it is
quite straightforward to calculate the ratio $R$ of (\ref{eq:Ren}). 
The results for a dijet case 
are shown by the dashed curves in Fig.~\ref{fig:upsd}  as a function of the rapidity $y_A$ of the 
dijet system.
The enhanced rescattering reduce the ratios  and lead to steeper $y_A$ distributions, as illustrated by the continuous curves. 
%
%
Perhaps the most informative probe of $S^2_{\rm en}$ is to observe the ratio $R$ for dijet production in the region $E_T \sim 15-30$ GeV. For example, for $E_T \sim$ 15 GeV we predict $S^2_{\rm en} \sim$ 0.25, 0.4 and 0.8 at $y_A=-2, ~0$ and $2$ respectively. 
\section{Conclusion}
 
The addition of forward proton detectors to LHC experiments will
add unique capabilities to the existing LHC experimental programme.
For certain BSM scenarios, the tagged-proton mode may even be the
Higgs discovery channel.
There is also a rich
QCD, electroweak, and 
more exotic physics, menu. 

The uncertainties in the prediction of the rate of a CEP process 
are potentially not small. Therefore, it is crucial
 to perform checks of the theoretical formalism
using processes that will be experimentally 
accessible in the first  runs of the LHC \cite{KMRearly}.

Most of the diffractive measurements described above can be performed, without detecting the forward protons, by taking advantage of the relatively low luminosity in the early LHC data runs. This allows the use of a veto trigger to select events with  a large rapidity gap(s). In this way we are able to probe the various individual components of the formalism used to predict the CEP cross sections. 

To summarise, the gap survival factor, $S^2$, caused by eikonal rescattering,
 may be studied as illustrated in Fig~\ref{fig:W3} and the possible enhanced, $S^2_{\rm en}$, contributions as shown in Figs.~\ref{fig:enh} and \ref{fig:upsd}. The relevant unintegrated gluon distribution, $f_g$, can be constrained by observing $\Upsilon$ production, see Fig.~\ref{fig:upsilon}, and the QCD radiative effect, $T$, may be checked by observing exclusive two- and three-jet events.

When the forward proton detectors are operating much more can be done. First, it is possible to measure directly the cross section $d^2\sigma_{\rm SD}/dtdM^2_X$ for single diffractive dissociation and also the cross section $d^2\sigma_{\rm DPE}/dy_1 dy_2$ for soft central diffractive production. These measurements will strongly constrain the models used to describe diffractive processes and the effects of soft rescattering. The 
recent predictions can be found in \cite{KMRnewsoft}. Next, a study of the transverse momentum distributions of both of the tagged protons, and the correlations between their momenta, is able to scan the proton optical density
\cite{KMRphoton,KMRtag}.

\section{Acknowledgements}
We thank Mike Albrow, Brian Cox, Albert De Roeck, Sasha Nikitenko, Andy Pilkington,
and Risto Orava for encouragement and for valuable advice.
 VAK is very grateful to Giorgio Bellettini,
 Giorgio Chiarelli and Mario Greco for the
 kind invitation and warm hospitality at  La Thuile.
This work was supported by
INTAS grant 05-103-7515 and by grant RFBR 07-02-00023.

%
\end{document}